\documentclass[aps,prl,twocolumn,superscriptaddress]{revtex4-1}

\usepackage{graphicx}
\usepackage{dcolumn}
\usepackage{bm}
\usepackage{amsmath}

\begin{document}

\preprint{APS/123-QED}

\title{A Stern-Gerlach experiment with light: separating photons by spin with the method of A. Fresnel}

\author{Oriol Arteaga}
\email{oarteaga@ub.edu}
\affiliation{Departament de F\'isica Aplicada, IN2UB, 08028 Barcelona, Spain.}
\affiliation{LPICM, \'{E}cole Polytechnique, Universit\'{e} Paris-Saclay, 91128 Palaiseau, France}

\author{Enric Garcia-Caurel}%
\affiliation{LPICM, \'{E}cole Polytechnique, Universit\'{e} Paris-Saclay, 91128 Palaiseau, France}

\author{Razvigor Ossikovski}
\affiliation{LPICM, \'{E}cole Polytechnique, Universit\'{e} Paris-Saclay, 91128 Palaiseau, France}

\date{\today}

\begin{abstract}

In 1822 A. Fresnel described an experiment to separate a beam of light into its right- and left- circular polarization components using chiral interfaces. Fresnel's experiment combined three crystalline quartz prisms of alternating handedness to achieve a visible macroscopic separation between the two circular components. Such quartz polyprisms were rather popular optical components in XIXth century but today remain as very little known optical devices. This work shows the analogy between Fresnel's experiment and Stern-Gerlach experiment from quantum mechanics since both experiments produce selective deflection of particles (photons in case of Fresnel's method) according to their spin angular momentum. We have studied a historical quartz polyprism with eight chiral interfaces producing a large spatial separation of light by spin. We have also constructed a modified Fresnel biprism to produce smaller separations and we have illustrated the process of weak measurement for light. The polarimetric analysis of a Fresnel polyprism reveals that it acts as a spin angular momentum analyzer.

\end{abstract}

\pacs{Valid PACS appear here}
\maketitle



One of the cornerstones of quantum mechanics is the Stern-Gerlach effect. In the original experiment from 1922 \cite{gerlach} an unpolarized beam of silver atoms is passed through a strong magnetic field gradient and splitting it into two polarized beams. In the experiment, particles with non-zero magnetic moment are deflected from straight path due to the magnetic field gradient. Historically, this experiment was decisive in convincing physicists of the reality of angular-momentum quantization in atomic-scale systems.

Particle spin is a truly quantum property and cannot be explained by classical physics. In fact, it was the discovery of particle spin that lead to the development of the modern theory of quantum mechanics. The photon also carries a quantity called spin angular momentum that does not depend on its frequency. The spin angular momentum of a particular photon is always either $+\hbar$ or $-\hbar$. These two possible helicities, called right-handed and left-handed, correspond to the two possible circular polarization states of the photon.

Because photons have no magnetic moment, they are not affected by the gradient of the magnetic field in the Stern-Gerlach apparatus, and the quantization of their angular momentum cannot be observed with this experiment. In some lectures about quantum mechanics \cite{feynman} an optical analog of a Stern-Gerlach experiment for photons is proposed using a calcite crystal, thanks to  the well-known phenomenon of double refraction that splits a beam of light into two orthogonally linearly polarized beams. This straightforward analogy is especially useful to understand sequential Stern-Gerlach experiments that involve a sequence of Stern-Gerlach apparatuses with each one being rotated with respect to the next one, because they can be explained by using the same mathematical model of states and measurements than that of a sequence of calcite polarizers; i.e. essentially, by using Malus law. However, the double refraction in calcite is not selecting or separating light according to the intrinsic angular momentum of photons. Another notorious difference is that, in the case of calcite double refraction, only one of the beams (namely, the extraordinary one) is deflected.

\begin{figure*}
\centering
  \includegraphics[width=15cm]{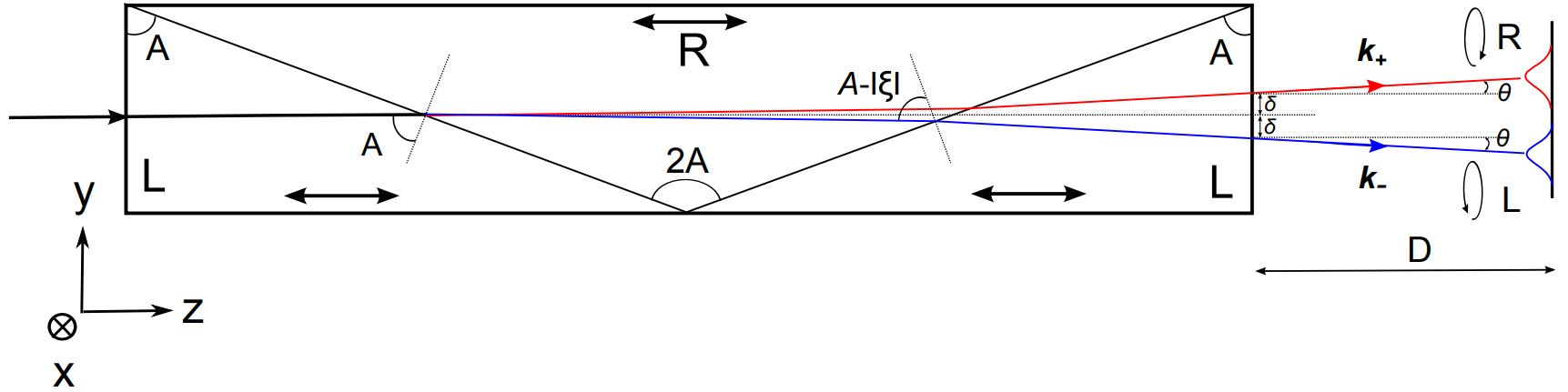}
\caption{Double refraction in a Fresnel triprism. The incident beam is split into right (R) and Left (L) circular waves. The double headed arrows inside the crystal indicate the direction of the optic axis, while $R$ and $L$ respectively denote a right- and a left-handed crystal. In Fresnel's original experiment $A=76^{\circ}$.}
  \label{trip}
\end{figure*}

This Letter studies a historical method proposed by Augustin J. Fresnel \cite{fresnel1} to produce circularly polarized light by selective deflection of photons according to their spin angular momentum and describes its analogy with the Stern-Gerlach experiment. In a memoir presented to the French academy of sciences in December 1822, Fresnel introduced for the first time the term \textit{circular polarization} and described a novel method to produce it. Previously, in 1817-1818 \cite{fresnel2,fresnel3}, he had already described what we know today as the Fresnel rhomb used to convert linearly polarized light into circularly polarized one by means of two total internal reflections introducing a $90^{\circ}$ phase difference. The second method, described by Fresnel in 1822, is much less known than the Fresnel rhomb one; it is based on the different speeds at which right- and left-circularly polarized waves propagate in a chiral medium. Specifically, Fresnel was considering natural light propagating along the optic axis of $\alpha$-quartz. In a bulk sample the difference in refractive indexes, an effect today known as optical rotation or circular birefringence, does not produce any physical separation of right- and left- circular polarizations but Fresnel described an ingenious experiment where the chiral interfaces between quartz prisms of alternating handedness cause double refraction of circularly polarized light (CPL), thus obtaining a macroscopic separation between the two circularly refracted beams. For this experiment Fresnel used a very acute quartz prism of angles $14^{\circ}-152^{\circ}-14^{\circ}$ placed between two other half-prisms of quartz of opposite handedness (of ``opposite species'' in Fresnel's terms). This optical design, shown in Fig. \ref{trip}, has been sometimes named as the \emph{Fresnel triprism} or, more generally, \emph{Fresnel polyprism}.

\begin{figure}
\centering
  \includegraphics[width=8.5cm]{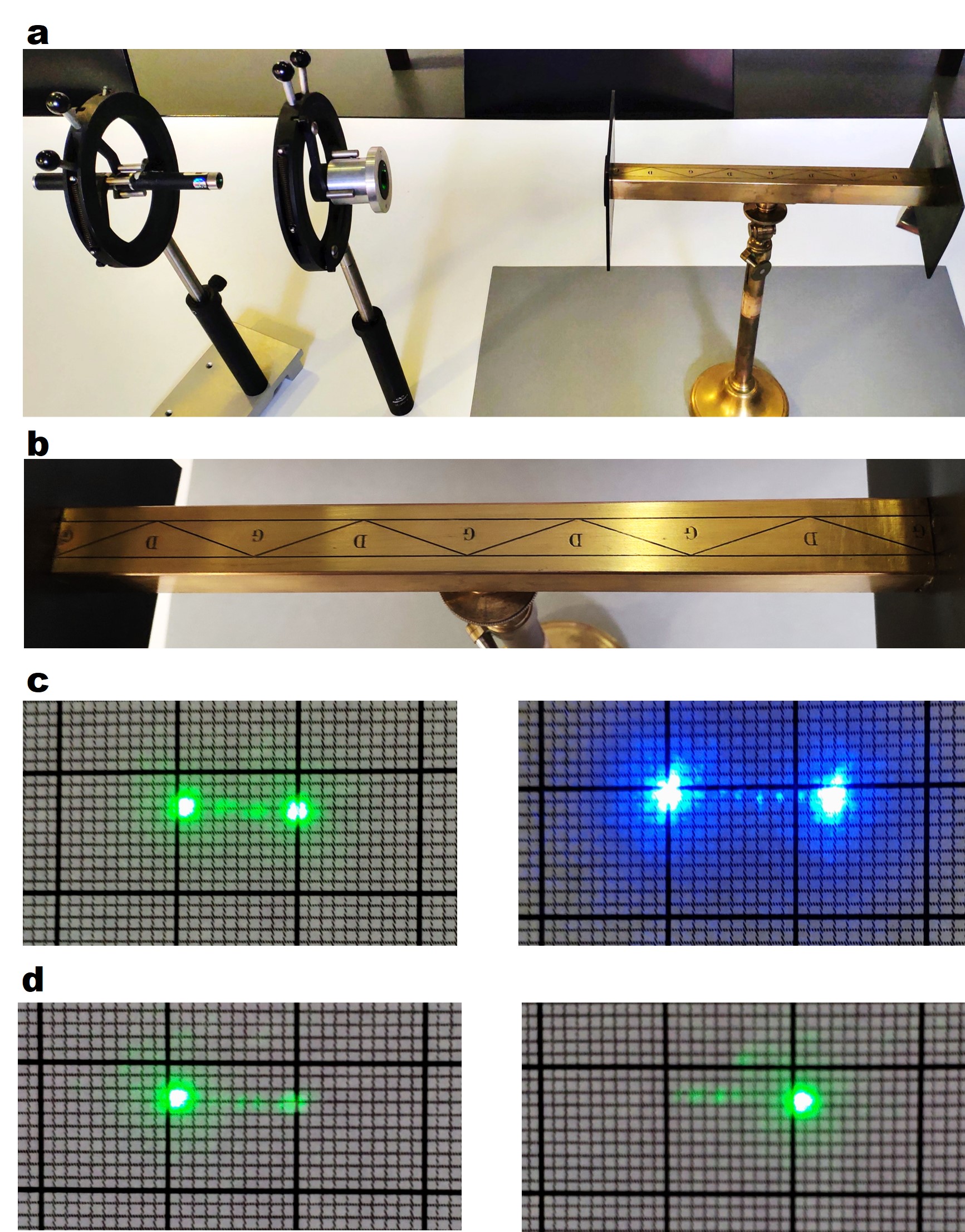}
\caption{a, Basic experiment showing the polyprism manufactured by H. Soleil in its brass mount, a laser pointer and a polarizer. b, Detail of the schematics engraved on the prism mount showing the internal design with 9 individual prisms. $D$ stands for the French ``droit'' (right) and $G$ for the French ``gauche'' (left). c, The spots projected on the screen for a 532 nm laser (left image) and a 405 nm laser (right image). In this experiment $D=208$ cm. d, screen projection when the laser light was pre-filtered to be right- and left-CPL respectively.   }
  \label{com}
\end{figure}

Fresnel polyprisms with alternating right- and left-quartz prisms were relatively popular optical devices in the XIXth century and, apparently, were even commercially available. However, they did not necessarily follow the original specifications for the prisms given by Fresnel (see Fig. \ref{com}). Nowadays, this method of producing circularly polarized light remains very little known --albeit it is mentioned in several books \cite{fowles,book1,sibook}-- and, to our knowledge, there are no reports of any recent fabrication of Fresnel polyprisms made of quartz. There are, however, some relatively recent works that have appraised Fresnel's method: in 1993 an old Fresnel prism was used as polarization interferometer to measure the circular birefringence of a sugar solution \cite{rodolfo} and, in 2006, Fresnel's method was used to study the optical rotation of a chiral liquid by using many prismatic cuvettes \cite{ghoshprl}. Curiously, Fresnel already anticipated in his memoir that his experiment could be also replicated using a chiral liquid such as turpentine instead of quartz. He estimated that to achieve effective separation with the materials he had at that time it would require 40 different cuvettes with liquids of alternating handedness. 

To our knowledge, the method of Fresnel, now almost 200 years old, has not been brought to the attention of investigators working on spin-controlled shaping of light. In recent years much interest has arisen in spin-dependent splitting of beams, i.e. situations in which the photon spin controls the propagation path and the intensity distribution of light, as it is considered to be a fundamental constituent of future spin-based photonics. Perhaps the most studied phenomenon is the so-called spin-Hall effect of light \cite{hosten,blio,yin,zhang,ling, ai2}, which is due to the interaction between photon spin and its orbital motion, so it is a consequence of light carrying both spin and orbital angular momentum. This effect is, for example, observed at a interfaces illuminated by finite size beams that are composed of many plane waves with slightly different wavevectors $\mathbf{k}$ and having slightly different planes of incidence. It manifests as a very small spin-dependent transverse shift upon reflection or refraction. This tiny shift is different from the more macroscopic effect discussed in this Letter, as the double refraction effect for CPL proposed by Fresnel is not directly related to the orbital angular momentum of light and can be readily explained in the plane wave picture.

A chiral medium such a quartz crystal can be described by the following Tellegen constitutive relations \cite{ossiyeh},
\begin{align}
\mathbf{D}=\bm{\varepsilon} \mathbf{E} +i\bm{\alpha} \mathbf{H}\\
\mathbf{B}=\bm{\mu} \mathbf{H} -i\bm{\alpha}^T \mathbf{E}
\end{align}
The permittivity and optical activity tensors of quartz have the following respective forms,
\begin{equation}\label{epsilon}
\bm{\varepsilon}=\mathrm{diag}\left(
  \begin{array}{ccc}
    \varepsilon_{11} & \varepsilon_{11} & \varepsilon_{33} \\
    \end{array}
\right)
\end{equation}
and 
\begin{equation}
\bm{\alpha}=\mathrm{diag}\left(
  \begin{array}{ccc}
    \alpha_{11} & \alpha_{11} & \alpha_{33} \\
    \end{array}
\right)
\end{equation}
The optical activity tensor was measured spectroscopically in Refs. \cite{recno,2mgequartz}. For propagation along the optic axis there are two forward  circularly polarized eigenmodes with different refractive indexes given by
\begin{equation}
n_{\pm}=\sqrt{\varepsilon_{11}}\mp\alpha_{11},
\end{equation}
where $n_+$ and $n_-$ respectively apply to right- and left-handed circular polarizations. Note that for right-handed quartz $n_->n_+$ because $\alpha_{11}$ is positive. The chiral parameter $\alpha_{11}$ is very small ($\sim 10^{-5}$) compared to the refractive index $\sqrt{\varepsilon_{11}}$. The normalized electric field eigenmodes are written as
\begin{equation}
\mathbf{E}_{\pm}=\frac{1}{\sqrt{2}}\left[\begin{array}{cc}
    1 & \pm i \\
    \end{array}\right]^T.
\end{equation}


For a left-handed crystal the permittivity tensor of Eq. \eqref{epsilon} remains the same but the optical activity tensor takes opposite values, so that $\alpha_{11}$ is now negative. Therefore, the refractive indexes for right- and left-handed crytas are related as
\begin{equation}
n^{LH}_{\pm}=n^{RH}_{\mp},
\end{equation}
where the superscripts $LH$ a $RH$ indicate left- and right-handed crystals respectively. In the following, we omit the superscripts $LH$ and $RH$ for simplicity and use a unique set of refractive indexes, namely those of a right-handed crystal. For example, if for right-handed quartz the refractive indexes are $n_+$ and $n_-$ for right- and left-CPL respectively, for a left handed quartz the values are reversed: right-CPL refracts with index $n_-$ and left-CPL with  $n_+$.

Plane wave reflection and refraction at an optical interface can be described by Snell's law and Fresnel relations. Because in an interface between  right- and left-handed quartz there is no discontinuity of the permittivity tensor, one can already anticipate that reflection and refraction phenomena will be mostly governed by the chiral parameter. In fact, the structure can be also regarded as a single medium with an inhomogeneous or stratified chirality. Notice that while in Stern-Gerlach experiment the deflection is produced by an inhomogeneous magnetic field, here the deflection of light is produced by an inhomogeneous chiral medium or by the interface between two chiral media of same nature but of opposite handedness. No deflection of light can take place without the presence of an interface.

Consider the first quartz-quartz interface shown in Fig. \ref{trip} between a left- a right-handed crystal. A right-handed circularly polarized wave obeys Snell's law,
\begin{equation}
n_-\sin{A}=n_+\sin{A'}.
\end{equation}

As in quartz $n_->n_+$ we can write $n_-=n_++\Delta$ where $\Delta=|n_R-n_L|=|2\alpha_{11}|$, therefore
\begin{equation}
\frac{\sin{A'}}{\sin{A}}=1+\frac{\Delta}{n_+},
\end{equation}
The refracted angle $A'$ can be rewritten as $A'=A+\xi$ where $\xi$ is the small angular deviation of the beam from the straight path due to the optical activity of quartz. Then 
\begin{equation}
\frac{\sin{(A+\epsilon})}{\sin{A}}\simeq1+\frac{\cos A}{\sin A}\epsilon=1+\frac{\Delta}{n_+}
\end{equation}
and 
\begin{equation}
|\xi|\simeq\Delta\tan A/n.
\end{equation}
where we have written $n_+\simeq n$, $n$ being the average refractive index; $n=(n_++n_-)/2$. A left-handed circularly polarized wave will deviate by the same angle (note that $n_-\simeq n_+\simeq n$) but in the opposite direction. 



In the second quartz-quartz interface of Fig. \ref{trip} the wave travels from a right-handed to a left-handed crystal. If, as before, we consider a right-handed circularly polarized wave, the Snell's law is
\begin{equation}
n_+\sin{(A-|\xi|)}=n_-\sin{A''}
\end{equation}
where we have taken into account that, due to the geometry of the triprism, the angle of incidence of the right-handed circular wave at this interface is $A-|\xi|$. Noticing that
\begin{equation}
n_+\sin{(A-|\xi|)}\simeq n_+\sin{A}(1-\Delta /n)\simeq (n_--2\Delta)\sin{A}
\end{equation}
where in the last step we have used that $n_+=n_--\Delta$. Rewriting the refraction angle $A''$ as $A''=A-\xi'$ where $\xi'$ is the small angular deviation of this beam with respect to the original incident beam, we find that 
\begin{equation}
|\xi'|\simeq 2\Delta\tan A/n,
\end{equation}
Finally, the beam goes through the last quartz-air interface which can be studied by applying again Snell's law,
\begin{equation}
n\sin{\xi'}=\sin{\theta}
\end{equation}
As all the involved angles are very small, $n\xi'\simeq\theta$, we obtain
\begin{equation}
\theta\simeq 2\Delta\tan A,
\end{equation}

If more prisms are added to the polyprism this formula can be generalized (approximating all small angles to first order) to 
\begin{equation}\label{defle}
\theta\simeq N\Delta\tan A
\end{equation}
where $N$ is the number of quartz-quartz interfaces of alternating handedness (in Fresnel triprism  $N=2$). The total angular separation between the right- and left-circularly polarized beams is $2\theta$. In this calculation we have assumed implicitly that light propagates always along the optic axis of quartz which, given the very small deflection angles, is a very reasonable approximation unless one considers a very large number of interfaces. If, instead of quartz, an isotropic chiral medium is considered, there would be no limitation to the number of successive interfaces one could use.




If we place a detector at a distance $D$ from the prism then the two beams will be separated by a distance $2|\Delta y(D)|=2(\delta+D\tan{\theta})$ where $\delta$ is the small separation at the exit facet of the device. The spin splitting $\delta$ at the exit of the polyprism depends on the angular deviation at each quartz-quartz interface and the pathlength through each prism. In the case of the Fresnel triprism considered in Fig. \ref{trip} it can be written as
\begin{equation}
\delta\simeq L\Delta \tan A/n
\end{equation}
where $L$ is the total length of the polyprism. The total transmitted field  at each point of a detection screen can be written as
\begin{equation}\label{basic}
\mathbf{E}(x,y,D)=\frac{1}{\sqrt{2}}[A_-e^{i\mathbf{k}_-\cdot\mathbf{r}}|-\rangle+A_+e^{i\mathbf{k}_+\cdot\mathbf{r}}|+\rangle]
\end{equation}
with
\begin{equation}
\mathbf{k}_{\pm}=\frac{2\pi n_0}{\lambda}[\pm\sin(\theta)\mathbf{\hat{y}}+\cos{\theta}\mathbf{\hat{z}}]
\end{equation}
where $n_0$ is the refractive index of the surrounding medium (in our case, the air). The vectors $|+\rangle$ and $|-\rangle$ are normalized Jones vectors for the right- and left-circularly polarized components respectively, 
\begin{equation}
|+\rangle=\frac{1}{\sqrt{2}}\left[\begin{array}{cc}
    1 &  i  \\
    \end{array}\right]^T,
\end{equation}
\begin{equation}
|-\rangle=\frac{1}{\sqrt{2}}\left[\begin{array}{cc}
    1 & - i  \\
    \end{array}\right]^T.
\end{equation}
$A_{\pm}$ are the amplitudes of the two beams at the detection point. As the two beams are separated along the $y$ coordinate we may write $A_{\pm}=A_0(x,y\pm \Delta y(D))$, where $A_0(x,y)$ can be considered as the amplitude of a Gaussian beam. The $y$ component of the wavevector $\mathbf{k}_{\pm}$ is the only one that depends on the spin, so we may explicitly refer to this component as $k_y$ and redefine it as $k_y=k_0\sigma\sin(\theta)$ where $\sigma$ is the spin, $\sigma=1$ for right-CPL and $\sigma=-1$ for left-CPL. 

An historical Fresnel polyprism was used to study experimentally Fresnel's method to split light by the spin angular momentum. This Fresnel polyprism (Fig. \ref{com}) was fabricated by Henri Soleil most likely between 1849 and 1872 and is currently on display in the museum of the \'{E}cole Polytechnique. This polyprism is not a triprism as originally described in Fresnel's memoir but rather contains nine different prisms of alternating handedness (i.e. eight interfaces between left- and right-handed quartz). We have found no other record of a Fresnel polyprism composed of such a large number of individual prisms. Despite its very old age, the polyprism seemed to be exceptionally well preserved. The experiments were carried out in the museum facilities, with the polyprism in its original mount and without any refurbishing of the optics. From the total length of the prism ($\sim 24$ cm) and the schematic drawing graved in the mount (see Fig. \ref{com}b) we inferred that the characteristic angle $A$ of each prism that composes the instrument is $\sim 73^{\circ}$.

Two different diode laser pointers (of respective wavelengths 532 nm and 405 nm) were used as light sources. After having traversed the polyprism, the light was projected on a screen covered with millimeter paper. Two well separated spots of equal intensities were effectively observed on the screen. We measured the separation between the two spots for both laser sources and for two different distances between the polyprism and the screen ($D=155$ cm and $D=208$ cm). As evidenced in Fig. \ref{com}c, 405-nm light featured larger separation between beams than 532-nm light. By adding a polarizer and a $\pm 45^{\circ}$ quarter-wave plate in front of the laser we repeated the experiment with light pre-filtered to a unique circular polarization state and, as expected, we obtained a single clear spot on the screen as displayed in Fig. \ref{com}d.

From the results of these experiments we could easily calculate the deflection angle $\theta$ and compare it with the predicted one given by Eq. \eqref{defle}. Table \ref{theo} compares the deflection angle values from the experiment with the calculation of Eq. \eqref{defle} assuming $N=8$, $A=73^{\circ}$, and the values of $\Delta=2\alpha_{11}$ obtained from the dispersion formula of $\alpha_{11}$ obtained in Ref. \cite{recno}. 

\setlength{\tabcolsep}{4pt}
\renewcommand{\arraystretch}{1.5}
\begin{table}[]
\begin{tabular}{@{}cccc@{}}
\hline
 Wavelength  & Measured $\theta$  & $\Delta$  & Calculated $\theta$ \\ \hline
 405 nm & $0.164^{\circ}\pm0.008^{\circ}$  & $1.090\cdot10^{-4}$  & $0.1633^{\circ}$   \\
 532 nm & $0.121^{\circ}\pm0.007^{\circ}$ & $7.920\cdot10^{-5}$  & $0.1187^{\circ}$   \\
 \hline
\end{tabular}
\caption{Measurements of the deflection angle $\theta$ for the historical polyprism in Fig. \ref{com} and comparison with the theoretical calculated values given by Eq. \eqref{defle} for the $\Delta$ values in Ref. \cite{recno} .}\label{theo}
\end{table}

The analogy between Fresnel's and Stern-Gerlach experiments also has a direct extension to the case of weak measurements. In 1988 Aharonov, Albert and Vaidman \cite{aharonov} suggested a modified Stern-Gerlach apparatus with a reduced magnetic field to measure the spin $z$ component of a particle with a well-defined spin state. Because of the weak magnetic field (wherefrom the term ``weak measurement'') the beams at the output are not be well separated but rather overlap. Theory showed that if a second Stern-Gerlach magnet with a strong magnetic field orthogonal to the first one is added (so that the spin along $x$ is measured), the separation of the two spin components would become much larger. Soon after an optical analogue was proposed \cite{duck} and experimentally realized \cite{ritchie}, by considering a coherent beam with Gaussian distribution that is first pre-selected in a linearly polarized state and then passed through a thin birefringent crystal producing a very small separation between the ordinary and extraordinary rays so that the two emerging beams overlap. If the composite emerging beam is post-selected with a polarizer orthogonal to the initial polarization state, a much larger separation of the two orthogonally polarized beams is observed, because the post-selection suppresses the overlapping part of the composite beam where the two orthogonally polarized rays merge. Such optical weak measurement amplification scheme has been used by several authors \cite{pfeiter, qiu, resch, zhang2,li}, most notably to study spin Hall effect that requires $\sim${\AA} spatial sensitivity \cite{hosten}.

\begin{figure}
\centering
  \includegraphics[width=7cm]{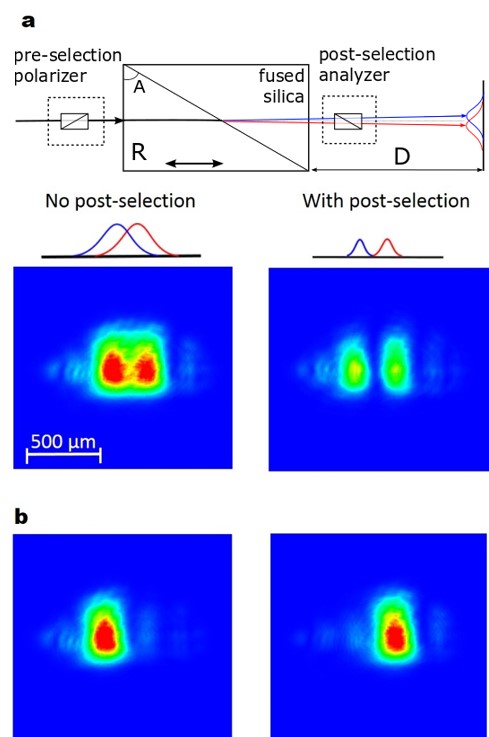}
\caption{Photometric analysis of the circular double refraction in a biprism. a, Basic schematics of the biprism and intensity profiles of the transmitted beam in measurements without and with post-selection respectively ($D=95$ cm). The post-selection was made by an analyzer orthogonal to the polarizer used for the pre-selection. b, Intensity profiles of the transmitted beam for pre-selection with right- and left-CPL (no post-selection).}
  \label{biprism}
\end{figure}

The weak measurement amplification method can be also applied to Fresnel prism. The analog of a Stern-Gerlach with weak magnetic field is a Fresnel prism with few interfaces and/or with a small angle $A$ leading, in accordance with Eq. \ref{defle}, to a very small deflection angle. In our experiment with the polyprism from Fig. \ref{com} the separation between the two circularly polarized states was so large that clearly no amplification is needed. Therefore, to illustrate the weak measurement amplification within Fresnel's method we set up a new experiment that offered a much smaller beam deflection. We used a single prism of right-handed $\alpha$-quartz in optical contact to another, achiral fused silica prism that can be considered as made of isotropic material with a refractive index $\sim\sqrt{\varepsilon_{11}}$. The angle $A$ of the prism was $63^{\circ}$. In this construction, shown in Fig. \ref{biprism}a, there is only one interface contributing to the beam splitting, and only one of the two interfaced media is chiral. The deflection angle is given by
\begin{equation}
\theta\simeq \frac{\Delta}{2}\tan A.
\end{equation}

In this experiment, a camera placed at a distance $D=95$ cm was used as a detector and the 405-nm laser diode pointer was used as a light source. A linear polarizer was placed in front of the biprism to pre-select a particular spin state. At this wavelength, the expected deflection for this biprism is $\theta \sim 0.0064^{\circ}$. When the light emerging from the biprism was directly measured by the camera we observed the intensity pattern of the bottom-left corner of Fig. \ref{biprism}a. Despite the small deflection, one can already anticipate from the image that splitting occurs. Note, however, that the splitting was not visible by simple eye inspection, so that, historically, Fresnel designed a sequence of prisms to amplify the effect. The beam splitting becomes much more evident when the emerging light is post-selected by an analyzer orthogonal to the initial polarizer, as shown in the bottom-right corner of Fig. \ref{biprism}a. The post-selection amplifies the beam separation but decreases the signal levels. Even if the optical activity of quartz is strong enough to produce a deflection measurable with modern photometric methods without any post-selection, this measurement illustrates how the weak measurement amplification scheme can be used in the optical Stern-Gerlach analog.

Next, we modified the experiment by adding a quarter-wave plate after the polarizer in order to illuminate the biprism with right- and left-CPL states. With this pre-selection, all incoming photons have the same spin angular momentum and follow a unique trajectory through the biprism, so that the camera detects a single spot, as expected (see Fig. \ref{biprism}b).

As a last experiment we studied the biprism in a Mueller matrix polarimeter and measured the Mueller matrix associated to each one the two possible paths followed by 405-nm light in the biprism. Details of the polarimeter used are given in Ref. \cite{4pem}. To allow the polarimeter discriminate between two very close emerging light spots we placed a pinhole of 20 $\mu m$ in front of the polarization state analyzer, at a distance of $\sim 54$ cm from the biprism. Then we adjusted precisely the position of the pinhole to select the portion of the emerging light that was analyzed. The normalized Mueller matrices measured at two slightly displaced positions of the pinhole were
\begin{equation}
\mathbf{M}_1=\left[
     \begin{array}{cccc}
       1 & -0.023& 0.014 & 0.968 \\
       0.011 & -0.005 & 0.036 & -0.012 \\
       0.012 & -0.021 & -0.012 & -0.006 \\
       0.984 & -0.006 & 0.012 & 0.992 \\
     \end{array}
   \right],
\end{equation}
\begin{equation}
\mathbf{M}_2=\left[
     \begin{array}{cccc}
       1 & 0.015 & 0.001 & -0.986 \\
       0.015 & -0.009 & 0.028 & -0.017 \\
       -0.016 & -0.031 & -0.022 & -0.012 \\
       -0.971 & -0.006 & 0.021 & 0.994 \\
     \end{array}
   \right].
\end{equation}
It is clear that $\mathbf{M}_1$ and $\mathbf{M}_2$ correspond, within the experimental noise (light levels were very low due to the very small pinhole), to the Mueller matrices of a right and left homogeneous circular polarizer respectively. Note that these are reciprocal forms of a circular polarizer, different from the Mueller matrix of a sequence of linear polarizer and a $\pm 45^{\circ}$ quarter-wave plate most often used to generate CPL. The Jones matrices corresponding to these homogeneous circular polarizers are
\begin{equation}
\mathbf{T}_+=\frac{1}{2}\left[
     \begin{array}{cc}
       1 & -i\\
       i & 1  \\
     \end{array}
   \right], \quad
\mathbf{T}_-=\frac{1}{2}\left[
     \begin{array}{cc}
       1 & i\\
       -i & 1  \\
     \end{array}
   \right].
\end{equation}
Matrices $\mathbf{T}_+$ and $\mathbf{T}_-$ are constructed from the outer products of the circular polarization Jones vectors and can be written as
\begin{equation}\label{plus}
\mathbf{T}_+=|+\rangle \langle+|=\frac{1}{2}(\mathbf{I}+\mathbf{S}_z),
\end{equation}
\begin{equation}\label{minus}
\mathbf{T}_-=|-\rangle \langle-|=\frac{1}{2}(\mathbf{I}-\mathbf{S}_z),
\end{equation}
where $\mathbf{I}$ is the 2$\times$2 identity matrix and $\mathbf{S}_z$ is the Pauli spin matrix
\begin{equation}
\mathbf{S}_z=\left[
     \begin{array}{cc}
       0 & -i\\
       i & 0  \\
     \end{array}
   \right].
\end{equation}
Subsequent multiplication of Eq. \eqref{plus} and \eqref{minus} on the right by, respectively, $|+\rangle$ and $|-\rangle$ generates the eigenvalue equations $\mathbf{S}_z|+\rangle=|+\rangle$ and $\mathbf{S}_z|-\rangle=-|-\rangle$, which show that $\mathbf{S}_z$ is the spin angular momentum operator for photons, with the two possible eigenvalues $\sigma=\pm 1$. Thus, the Mueller matrix measurements we have carried out verify experimentally that Fresnel polyprisms, like Stern-Gerlach apparatus, perform spin measurement.

Although the notion of circular polarization was introduced by Fresnel almost 200 years ago, and for more than a century it is has been known that light possesses both linear and angular momenta along the direction of propagation, only very recently there has been exceptional increase in the number of publications and applications that exploit the spin angular momentum of light. This development is strongly connected to recent advances in nanophotonics and plasmonics, as well as to recent studies of tiny optical spin-Hall effects caused by spin-orbit interactions. However, the more macroscopic spin-dependent deflection of light described by Fresnel has not yet been exploited for modern spin-dependent transportation phenomena. In this context, promising fields for the application of Fresnel's spin-dependent transportation appear to be on-chip and interchip optical circuitry and optical quantum computing \cite{applications}. Therefore, Fresnel's method could grant access to selective directive and spin-sensitive addressing or to reading of guided modes and quantum states. More generally, the possibilities of Fresnel's method as a generator or a detector of circularly polarized light seem to be not yet fully exploited.

In summary, we have studied an optical analog of the original Stern-Gerlach experiment  producing a macroscopic separation between the spin states of a light beam. Very remarkably, Fresnel described this spin splitting experiment exactly one century before the actual Stern-Gerlach experiment was published.






\section*{Acknowledgments}

This work was partially funded by Ministerio de Economia y Competitividad (EUIN2017-88598) and European Commission (Polarsense, MSCA-IF-2017-793774). Authors are grateful to Marie-Christine Thooris and Olivier Azzona, curators of the museum of the \'{E}cole Polytechnique, for allowing us to use the Fresnel polyprism by H. Soleil in the experiments.

\bibliography{bibliography}

\end{document}